\documentclass{Interspeech2024}

\usepackage{multirow,enumitem,adjustbox}
\usepackage{booktabs,threeparttable,makecell}
\usepackage{subcaption}
\usepackage{url}

\interspeechcameraready

\title{Generating Speakers by Prompting Listener Impressions\\ for Pre-trained Multi-Speaker Text-to-Speech Systems
}

\name[affiliation={1,2}]{Zhengyang}{Chen}
\name[affiliation={2}]{Xuechen}{Liu}
\name[affiliation={3}]{Erica}{Cooper}
\name[affiliation={2}]{Junichi}{Yamagishi}
\name[affiliation={1}]{Yanmin}{Qian}

\address{
  $^1$AudioCC Lab, CS Dept, Shanghai Jiao Tong University, Shanghai, China\\
  $^2$National Institute of Informatics, Tokyo, Japan\\
  $^3$National Institute Of Information And Communications Technology, Kyoto, Japan
  }
\email{zhengyang.chen@sjtu.edu.cn, jyamagis@nii.ac.jp}


\begin{document}

\maketitle

\begin{abstract}
This paper proposes a speech synthesis system that allows users to specify and control the acoustic characteristics of a speaker by means of prompts describing the speaker's traits of synthesized speech.
Unlike previous approaches, our method utilizes listener impressions to construct prompts, which are easier to collect and align more naturally with everyday descriptions of speaker traits. We adopt the Low-rank Adaptation (LoRA) technique to swiftly tailor a pre-trained language model to our needs, facilitating the extraction of speaker-related traits from the prompt text. Besides, different from other prompt-driven text-to-speech (TTS) systems,  we separate the prompt-to-speaker module from the multi-speaker TTS system, enhancing system flexibility and compatibility with various pre-trained multi-speaker TTS systems. Moreover, for the prompt-to-speaker characteristic module, we also compared the discriminative method and flow-matching based generative method and we found that combining both methods can help the system simultaneously capture speaker-related information from prompts better and generate speech with higher fidelity.
\end{abstract}

\section{Introduction}

Multi-speaker text-to-speech systems \cite{gibiansky2017deep,cooper2020zero,casanova2022yourtts} aim to synthesize natural speech conditioned on the specific content text and target speaker information. The speaker information can be provided by speaker ID, reference speech, or encoded speaker embedding. However, the available speaker ID must be used in the training process and the reference speech could be hard to find in a short period if we want to create some unseen voices.
Besides, providing reference speech may not be user-friendly for some ordinary users.

Natural language serves as the most intuitive and comprehensive medium for humans to communicate information. Recent research endeavors have aimed at harnessing this capability within text-to-speech (TTS) systems by controlling speaker-related attributes through textual descriptions, commonly referred to as prompts. Studies such as those by Guo et al. \cite{guo2023prompttts}, Leng et al. \cite{leng2023prompttts}, Liu et al. \cite{liu2023promptstyle}, and Yang et al. \cite{yang2023instructtts} mainly explore the manipulation of style-related attributes via text prompts. Conversely, Zhang et al. \cite{zhang2023promptspeaker} investigated the modulation of speaker identity information. Extending this domain, Shimizu et al. \cite{shimizu2023prompttts++} used prompts to concurrently modulate both style and speaker identity attributes.


Despite notable advancements in prompt-driven text-to-speech (TTS) technology, several persistent challenges merit further investigation. The authors in~\cite{guo2023prompttts,leng2023prompttts} have trained their systems using datasets with paired speech and prompt descriptions. However, acquiring TTS training data is much easier than procuring prompt-specific data \cite{yang2023instructtts,zhang2023promptspeaker}. This discrepancy suggests that decoupling the TTS model from the prompt-modulation model may be advantageous. Typically, the pre-trained language models (LM) used for encoding prompt information are developed using general-purpose datasets. As such, it may not suffice to merely integrate basic modules \cite{yang2023instructtts,zhang2023promptspeaker} atop these LMs to tailor them for TTS applications. 
Meanwhile, the methods for collecting prompt data can be categorized into two main approaches: deriving statistical signal processing measures~\cite{guo2023prompttts,leng2023prompttts}, such as pitch and speed, from larger datasets automatically; or directly collecting small-scale prompts manually~\cite{yang2023instructtts,zhang2023promptspeaker}, which involves a more curated and thus potentially less scalable process.
Identifying more effective strategies for gathering prompt data remains a crucial area for exploration.

We propose generating the prompts from listener impression scores, which can be more easily collected than the complete prompt descriptions and align more closely with natural descriptions of voice in daily conversations compared with the signal processing statistics-based prompts. Furthermore, we address the challenge of pre-trained LMs, which are typically trained on general datasets that may not effectively capture nuances related to speaker identity and speaking styles. To this end, we use a low-rank adaptation strategy (LoRA) \cite{LoRA}, adapting the pre-trained LM to better suit our specific requirements. Our experimental results underscore the significance of the LoRA module in enhancing overall performance. Additionally, different from the previous works~\cite{guo2023prompttts,leng2023prompttts}, we propose a modular design for the prompt-based TTS system, decoupling the prompt-to-speaker module from the TTS system. This separation increases the system's flexibility, allowing for seamless integration with various multi-speaker TTS frameworks. When mapping the prompt to another modality, researchers have used either a discriminative method~\cite{liu2023promptstyle,yang2023instructtts,clip,clap,huang2022mulan} or generative method~\cite{zhang2023promptspeaker}. Our findings indicate that each method offers distinct benefits, and a hybrid approach that combines both methods yields further enhancements.



\section{Prompt-driven Speaker Generation}


\subsection{System Overview}
\begin{figure}
    \centering
    \includegraphics[width=\linewidth]{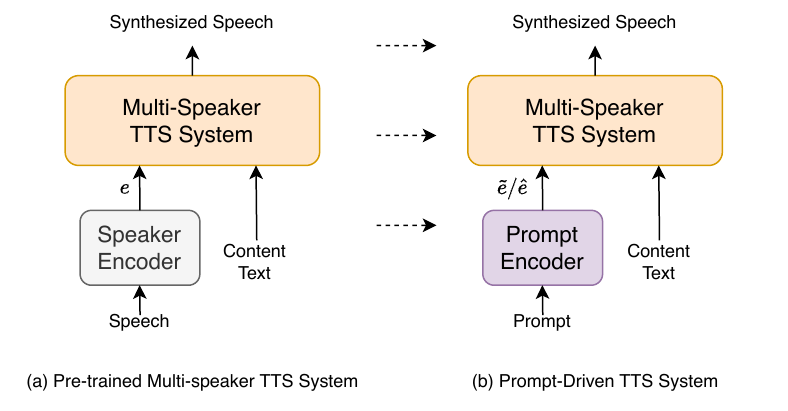}
    \caption{Overview of our system. 
    $\Tilde{e}$ and $\hat{e}$ are two types of outputs of the prompt encoder. 
    Refer to Figure \ref{fig:prompt2emb} for more details.
    }
    \label{fig:overview}
\end{figure}

As shown in Figure \ref{fig:overview}, our methodology extends the text-to-speech (TTS) task by utilizing both content text and the prompt from listener impressions as inputs. The content text controls the linguistic aspects of the generated speech, while the prompt from listener impressions modulates the speaker's characteristics.
We detail the process of prompt construction in section \ref{ssec:datasets}.
Our approach begins with pre-training a Variational Inference with adversarial learning for end-to-end Text-to-Speech (VITS) system~\cite{kim2021conditional},  which is modified in our experiment to condition on speaker embeddings $e$ derived from an external speaker encoder. Furthermore, we replaced the original speaker encoder with a prompt encoder. This modification necessitates that the prompt encoder is capable of accurately mapping prompts to their respective speaker embeddings, thereby enabling the precise control of speaker characteristics through textual prompts. 



In the following sections, we introduce two methods to map the prompt text to speaker embedding, the discriminative method and the generative method. In the discriminative method, the speaker embedding is deterministically determined by the prompt, which is widely used in previous multi-modal linking models~\cite{clip,clap,huang2022mulan}. Besides, we also propose to use the generative flow-matching~\cite{DBLP:conf/iclr/LipmanCBNL23} model to learn the distribution of the speaker embeddings conditioned on the prompt.

\label{ssec:system_overview}
 
\subsection{Discriminative Method}
\label{ssec:discriminative}

\begin{figure}[ht!]
    \centering
    \includegraphics[width=\linewidth]{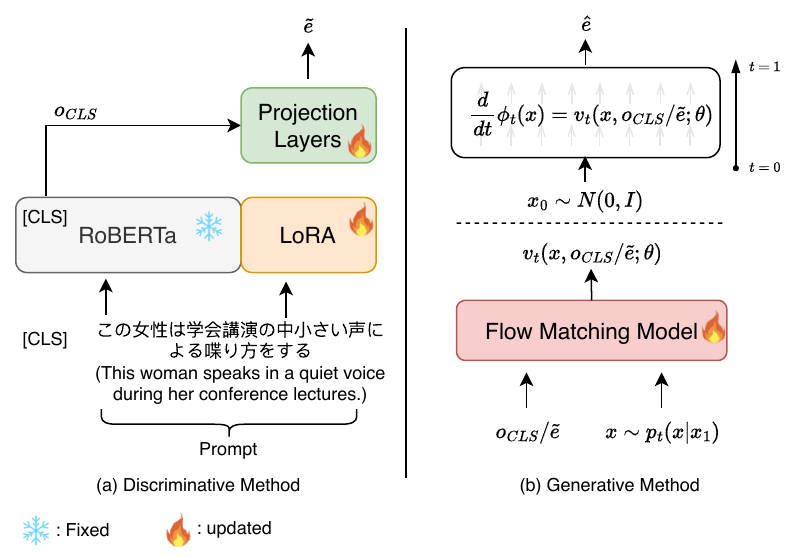}
    \caption{The Prompt Encoder Design.}
    \label{fig:prompt2emb}
\end{figure}

In this section, we introduce a discriminative model to map the text prompt to speaker embedding. Unlike other multi-modal linking models, e.g. CLIP \cite{clip} and CLAP \cite{clap}, we update only the text prompt encoder here, which enables our model to be easily adapted to any pre-trained multi-speaker text-to-speech system. As depicted in Figure \ref{fig:prompt2emb}(a), each text prompt is initially appended with a $[\textit{CLS}]$ token. This modified prompt is then processed by RoBERTa~\cite{liu2019roberta}\footnote{\url{https://huggingface.co/nlp-waseda/roberta-base-japanese-with-auto-jumanpp}}, for which the output at the $[\textit{CLS}]$ token, denoted as $o_{CLS}$, encapsulates the comprehensive information of the text prompt. Finally, $o_{CLS} \in \mathbb{R}^{d'}$ is fed into another projection module to obtain the predicted speaker embedding $\Tilde{e} \in \mathbb{R}^{d}$. Considering that many speaker recognition systems optimize the speaker embedding in the hyper-sphere space~\cite{huang2018angular,DBLP:conf/interspeech/LiuHL19a}, we update the discriminative model by simultaneously minimizing the L2 distance and maximizing the cosine similarity between $\Tilde{e}$ and the ground truth embedding $e$. The loss function is formulated as follows:
\begin{equation}
    \mathcal{L} = \Vert \Tilde{e} - e\Vert^2 + (1 - \textit{cosine\_similarity}(\Tilde{e}, e))
\end{equation}
We also explore using the LoRA~\cite{LoRA} in Figure \ref{fig:prompt2emb}(a) module to enhance the RoBERTa for our task and we consider the RoBERTa without LoRA as our baseline in our experiment.

\subsection{Generative Method based on Flow Matching}
\label{ssec:generative}

Although discriminative multi-modal linking methods have shown commendable performance in downstream tasks, e.g. prompt-driven speech generation~\cite{yang2023instructtts}, image generation~\cite{ramesh2022hierarchical} and audio generation~\cite{DBLP:conf/icml/LiuCYMLM0P23}, the relationship between text prompts and speaker embeddings is not strictly one-to-one. A single prompt can often describe different speakers, highlighting a complex one-to-many mapping challenge. To address this inherent complexity, we propose the adoption of a Flow Matching (FM) based generative model \cite{DBLP:conf/iclr/LipmanCBNL23} for generating speaker embeddings from text prompts. 

\subsubsection{Flow Matching Algorithm}
Modeling the distribution of data points $x_1 \in \mathbb{R}^d$ sampled from an unknown distribution $q(x_1)$ using deep learning techniques presents significant challenges. The generative model is always designed to learn the transformation from a simple prior distribution $p_0$ (e.g., a Gaussian distribution) to a target distribution $p_1 \approx q$. The flow matching algorithm \cite{DBLP:conf/iclr/LipmanCBNL23} is proposed to construct a continuous flow $\phi_t:\mathbb{R}^d \rightarrow \mathbb{R}^d, t \in [0, 1]$ for transforming the prior distribution into the target distribution by regressing the vector field $u_t \in \mathbb{R}^d$. The relationship between the flow and vector field is formulated using an ordinary differential equation (ODE):
\begin{equation}
    \frac{d}{d t} \phi_t(x)=u_t\left(\phi_t(x)\right)
\end{equation}

Thus, if we can approximate $u_t$ using a neural network, we can construct the flow path. However, given the absence of a closed-form expression for $u_t$, we cannot approximate it directly. Lipman et al.~\cite{DBLP:conf/iclr/LipmanCBNL23} propose utilizing a conditional vector field $u_t(x|x_1)$ to replace the original vector field $u_t$, leading to the Conditional Flow Matching (CFM) objective:
\begin{equation}
\label{eq:CFM}
\mathcal{L}_{\mathrm{CFM}}(\theta)=\mathbb{E}_{t, q(x_1), p_t(x|x_1)}\Vert v_t(x, \theta)-u_t(x|x_1)\Vert^2
\end{equation}
where $p_t(x|x_1)$ denotes the probability density function conditioned on $x_1$ at time $t$, and $v_t(x, \theta)$ is the neural network we used to approximate $u_t(x|x_1)$. The authors in \cite{DBLP:conf/iclr/LipmanCBNL23} also prove that approximating $u_t(x|x_1)$ is equivalent to approximating $u_t$.


To define the path of the flow, we utilize the optimal transport (OT) path as described in \cite{DBLP:conf/iclr/LipmanCBNL23}, where $p_t(x | x_1) = \mathcal{N}(x | t x_1, (1 - (1 - \sigma_{\min}) t)^2 I)$ and $u_t(x | x_1) = (x_1 - (1 - \sigma_{\min}) x) / (1 - (1 - \sigma_{\min}) t)$. Here, $\sigma_{\min}$ is a scalar marginally above zero.

\subsubsection{Generate Speaker Representation based on Flow Matching}
\label{sssec:FM_prompt_encoder}
In this study, our objective is to generate speaker embeddings that are conditioned on the prompt from listener impressions. Illustrated in Figure \ref{fig:prompt2emb} and following the approach described in Section \ref{ssec:discriminative}, we initially process the prompt through the RoBERTa model with a LoRA module, yielding the output $o_{CLS}$. To condition the CFM model on the prompt, we reformulate the approximated vector field in equation \ref{eq:CFM} to $v_t(x, o_{CLS}; \theta)$. We can also condition the FM model on the output of the discriminative model to build a two-stage system, and the vector field is formulated as  $v_t(x, \Tilde{e}; \theta)$.  During the inference phase, speaker embeddings $\hat{e}$ are generated by integrating the ODE function from $t=0$ to $t=1$:
\begin{equation}
    \frac{d}{d t} \phi_t(x)=v_t(x, o_{CLS}/\Tilde{e}; \theta); \phi_0(x) = x_0 \sim N(0, I)
\end{equation}
To balance the generative fidelity and time consumption, we set the ODE step to 32 in our experiment.

\section{Experiment Setup}

\subsection{Dataset and Prompt Construction}
\label{ssec:datasets}


\begin{figure}[t]
    \centering
    \includegraphics[width=\linewidth]{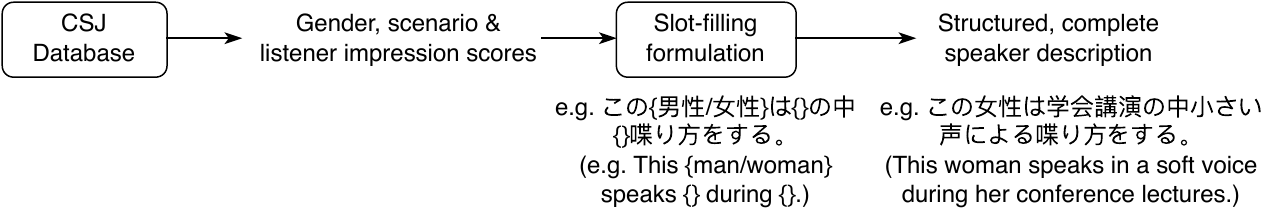}
    \caption{\textbf{Prompt construction pipeline from listener impressions.} The prompt is created using slot-filling techniques, with impression phrases filling the two slots indicated by brackets.}
    \label{fig:prompt_construction}
\vspace{-10pt}
\end{figure}

In our work, we leverage the Corpus of Spontaneous Japanese (CSJ) \cite{csj_corpus} dataset and follow the dataset partition in \cite{csj_official_asr2015}, resulting in 2,672 and 30 speakers for training and evaluation, respectively. Meanwhile, we isolated 200 utterances from 20 speakers in the trainset to form the held-out validation dataset, which is not used for model training. 
Even though the CSJ dataset has its own transcripts, there is no punctuation, which is important for the TTS system. To generate transcripts with punctuation for the CSJ dataset, we pre-process the CSJ dataset by leveraging the small-version pre-trained Whisper \cite{radford2023robust} model. 

The CSJ dataset also provides listener impression test scores for speaker characteristics. According to the description available at the website\footnote{\url{https://clrd.ninjal.ac.jp/csj/manu-f/impression.pdf}}, it comprises both binary inquiries (e.g., high/low pitch, old/young) and rank-order queries on a five-point scale (e.g., speaking speed, demeanor), resulting in 26 questions in total. Each of the scores for each question can be reformulated as a phrase describing speaker impression. The process of building descriptions from the listener impression test scores are illustrated in Figure \ref{fig:prompt_construction}. 

\subsection{Model Configuration}
\label{ssec:model_config}
In our experiment, we use the pre-trained r-vector (ResNet34) from the wespeaker\footnote{\url{https://github.com/wenet-e2e/wespeaker/blob/master/docs/pretrained.md}} \cite{wang2023wespeaker} as the speaker encoder for the multi-speaker text-to-speech system. We follow the VITS implementation in this repository\footnote{\url{https://github.com/jaywalnut310/vits}} to leverage the external speaker embedding.
For the prompt encoder in our experiment, we implement the LoRA module following the AdapterHub\footnote{\url{https://github.com/adapter-hub/adapters}}~\cite{pfeiffer2020AdapterHub} toolkit and set the LoRA rank to 8. We implement the Projection module introduced in section \ref{ssec:discriminative} as 4-layer linear layers. We also design the Flow Matching model introduced in Section \ref{ssec:generative} in the same way as the Projection module. When combining the discriminative method with the flow-matching based generative method introduced in section \ref{sssec:FM_prompt_encoder}, we simply stack the Flow Matching model in Figure \ref{fig:prompt2emb}(b) on the Projection model in \ref{fig:prompt2emb}(a).

In our experiment, we first pre-train the multi-speaker TTS system on the CSJ training set, during which the speaker encoder is fixed. Then, we train the prompt encoder based on the speaker embeddings and prompts introduced in Sections \ref{ssec:datasets} and \ref{ssec:model_config}. During inference, we simply replace the speaker encoder in the multi-speaker TTS system with the prompt encoder to enable prompt-driven text-to-speech.

\subsection{Evaluation Metric}
\label{ssec:eval_metric}

\subsubsection{Objective Evaluation}
\label{ssec:objective_eval}
Due to the one-to-many mapping nature of the prompt-to-speaker generation task introduced in Section \ref{ssec:generative}, we do not have an exact ground-truth reference for each generated speaker embedding and generated audio sample. Here, we borrow the reference-free evaluation metric, Fréchet Audio Distance (FAD)~\cite{kilgour2018fr}, for our experiment. In the FAD evaluation, we randomly select 5,000 audio samples from training set as the background speech set. Utilizing the Encodec~\cite{defossez2022high} model from the fadtk toolkit\footnote{\url{https://github.com/microsoft/fadtk}} \cite{fadtk}, we extract embeddings from both this background set and the synthesized speech generated from prompts in the CSJ evaluation set. Then, FAD scores are calculated based on the extracted embeddings. A lower FAD score means that the synthesized speech has a similar distribution to the background speech set, indicating better audio fidelity. 


\begin{table*}[ht!]
\footnotesize
\centering
\caption{\textbf{Spearman Rank Correlation Coefficient (SRCC) between MOS scores from reference and synthesized speech}. }

\begin{adjustbox}{width=.99\textwidth,center}
\begin{threeparttable}
\begin{tabular}{ll|ccccccccc|c}
\toprule 
\multirow{2}{*}{Scenario} & \multirow{2}{*}{System} & \multicolumn{10}{c}{Speaker Attribute}\\
\cline{3-12}
& & expressiveness & confidence & relaxation & voice\_depth & age & energy & pitch & speed & clarity & Avg \\
\hline
\multirow{4}{*}{Seen} & Discriminative (w/o LoRA) & \underline{0.72} & 0.53 & 0.48 & \underline{0.75} & \underline{0.86} & \underline{0.71} & \underline{0.89} & \underline{0.89} & 0.23 & 0.67 \\
& Discriminative (w/ LoRA) & \underline{0.71} & \underline{0.69} & 0.65 & \underline{0.83} & \underline{0.90} & \underline{0.76} & \underline{0.94} & \underline{0.85} & 0.37 & \textbf{0.74} \\
& Flow-Matching (w/ LoRA) & \underline{0.68} & 0.53 & 0.66 & \underline{0.76} & \underline{0.79} & 0.50 & \underline{0.86} & 0.38 & 0.22 & 0.60 \\
& Discriminative + Flow-Matching & \underline{0.74} & \underline{0.71} & \underline{0.75} & \underline{0.87} & \underline{0.96} & \underline{0.72} & \underline{0.90} & \underline{0.68} & 0.35 & \textbf{0.74} \\
\midrule
\multirow{4}{*}{Unseen} & Discriminative (w/o LoRA) & \phantom{-}0.04 & 0.05 & 0.46 & 0.38 & 0.67 & 0.29 & \underline{0.73} & 0.57 & -0.37 & 0.31 \\
& Discriminative (w/ LoRA) & \phantom{-}0.54 & 0.38 & 0.49 & 0.48 & \underline{0.77} & 0.25 & \underline{0.81} & 0.36 & 0.41 & \textbf{0.50} \\
& Flow-Matching (w/ LoRA) & -0.10 & 0.12 & 0.32 & 0.42 & \underline{0.82} & 0.39 & \underline{0.74} & 0.14 & 0.21 & 0.34 \\
& Discriminative + Flow-Matching & \phantom{-}0.36 & 0.08 & 0.49 & 0.35 & \underline{0.74} & 0.34 & \underline{0.75} & 0.37 & 0.20 & 0.41 \\
\bottomrule

\end{tabular}
\begin{tablenotes}\footnotesize
\item \underline{underline}: The statistical significance (p-value) is less than 0.001, indicating the MOS scores of synthetic speech are significantly correlated with the MOS scores of reference speech.
\end{tablenotes}
\end{threeparttable}
\label{table:srcc_res}
\end{adjustbox}
\end{table*}

\subsubsection{Subjective Evaluation}
\label{ssec:subjective_eval}
We conducted a listening test and recruited 100 native Japanese listeners to evaluate both the synthesis quality and the ability of the synthesis systems to produce speech that correctly reflects the speaker attributes described in the prompt. 
We first select 100 utterances (10 male and 10 female each, 5 utterances for each speaker) from the CSJ evaluation (unseen speaker) and held-out validation set (seen speaker), respectively, as the natural speech reference set. Then, we use the prompts and content text according to these 200 utterances to generate 200 utterances using each of the four systems.  We first
asked listeners to rate the samples on a scale of 1-5 for overall naturalness. We also asked listeners to give their impressions about nine different speaker attributes on a 5-point rating scale. 
For each speaker attribute, each sample from the reference set and the synthesized audio is rated 8 times by different raters. Since each speaker corresponds to 5 utterances, there are 40 MOS scores per speaker from the same attribute. Then we average the 40 MOS scores for each speaker to remove the randomness.

\section{Results}
\subsection{Audio Fidelity and Naturalness Evaluation}
\label{ssec:fad_res}
We employ FAD score and naturalness MOS, detailed in Section \ref{ssec:eval_metric}, to assess the fidelity and naturalness of synthesized speech from both objective and subjective perspectives. Results in Table \ref{table:fad_score} reveal the indispensable role of the LoRA module in enhancing speech synthesis, corroborating our hypothesis that merely augmenting the language model with additional layers is insufficient for this task. Furthermore, we demonstrate that our novel approach of generating speaker embeddings through the generative flow-matching model surpasses discriminative methods in terms of speech fidelity and naturalness. Notably, the combination of discriminative and generative techniques yields further improvement in the fidelity of synthesized speech.



\begin{table}[t]
\footnotesize
\centering
\caption{\textbf{FAD score and Naturalness MOS results on the CSJ evaluation set.}
}
\begin{adjustbox}{width=.45\textwidth,center}
\begin{threeparttable}
\begin{tabular}{l|cc}
\toprule
System & FAD Score & Naturalness MOS \\
\hline
ground-truth & - & 4.06 $\pm$ 0.25 \\

Discriminative (w/o LoRA) &  11.217 & 3.15 $\pm$ 0.25 \\
Discriminative (w/ LoRA) &  \phantom{0}5.244 & 3.45 $\pm$ 0.19 \\
Flow-Matching  (w/ LoRA) & \phantom{0}3.559  & 3.52 $\pm$ 0.26 \\
Discriminative + Flow-Matching & \phantom{0}3.126 & 3.50 $\pm$ 0.24 \\


\bottomrule
\end{tabular}
\end{threeparttable}
\label{table:fad_score}
\end{adjustbox}
\end{table}



\subsection{Speaker information relevance between synthesized speech and prompt}

In Section \ref{ssec:subjective_eval}, we evaluate our systems in both seen and unseen speaker scenarios by collecting 20 Mean Opinion Score (MOS) ratings (corresponds to 20 speakers) for each system regarding a specific speaker attribute. We calculate the Spearman Rank Correlation Coefficient (SRCC)~\cite{sedgwick2014spearman} between the MOS scores from synthesized speech and reference speech and list the results in Table \ref{table:srcc_res}. 

\begin{figure}[t]
\begin{subfigure}{.23\textwidth}
  \centering
  \includegraphics[width=1.0\linewidth]{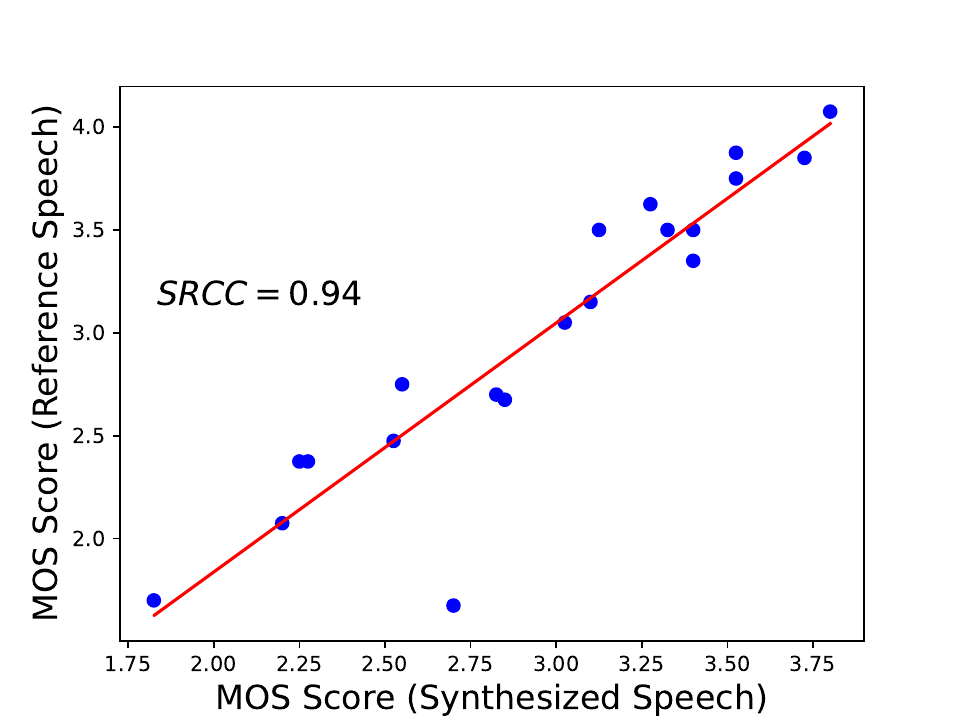}  
  \caption{Pitch MOS score.}
  \label{fig:seen_pitch}
\end{subfigure}
\begin{subfigure}{.23\textwidth}
  \centering
  \includegraphics[width=1.0\linewidth]{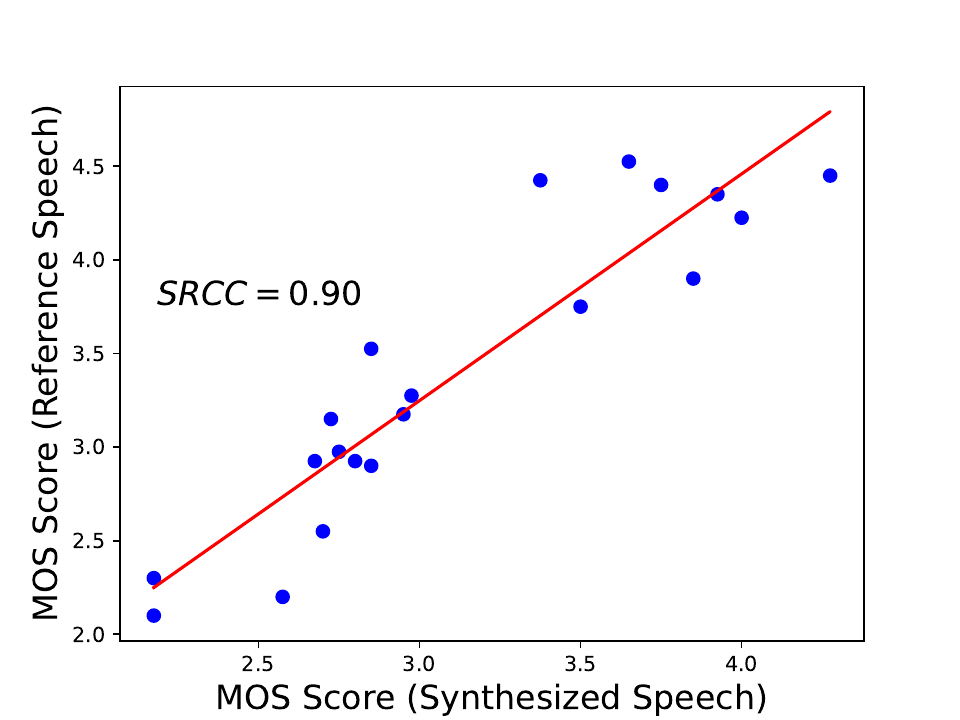}  
  \caption{Age MOS score.}
  \label{fig:seen_age}
\end{subfigure}

\begin{subfigure}{.23\textwidth}
  \centering
  \includegraphics[width=1.0\linewidth]{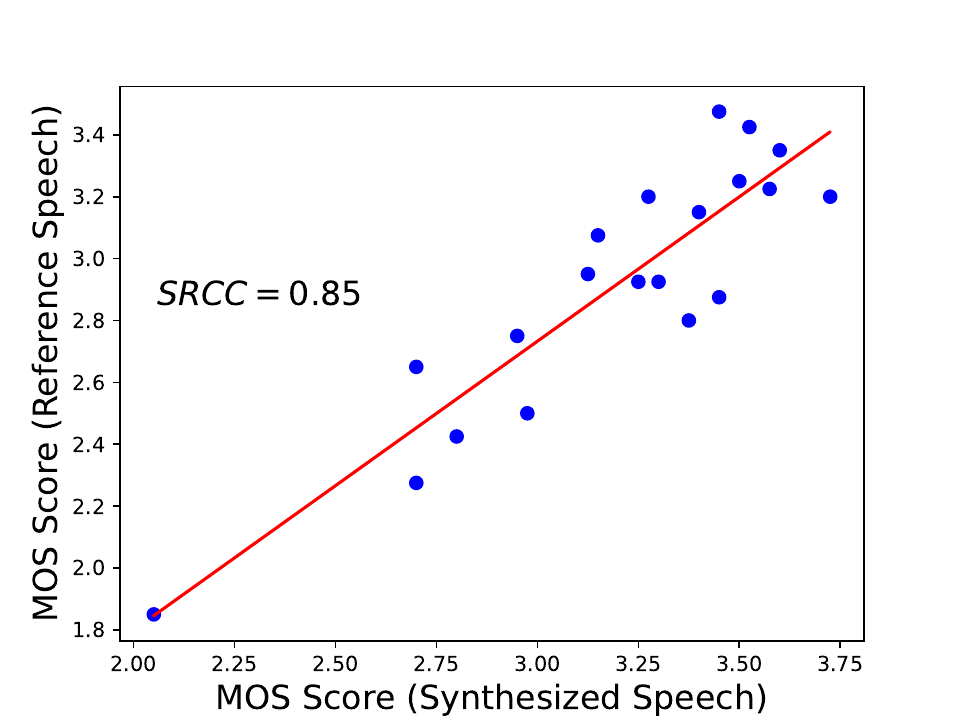}  
  \caption{Speed MOS score.}
  \label{fig:seen_speed}
\end{subfigure}
\begin{subfigure}{.23\textwidth}
  \centering
  \includegraphics[width=1.0\linewidth]{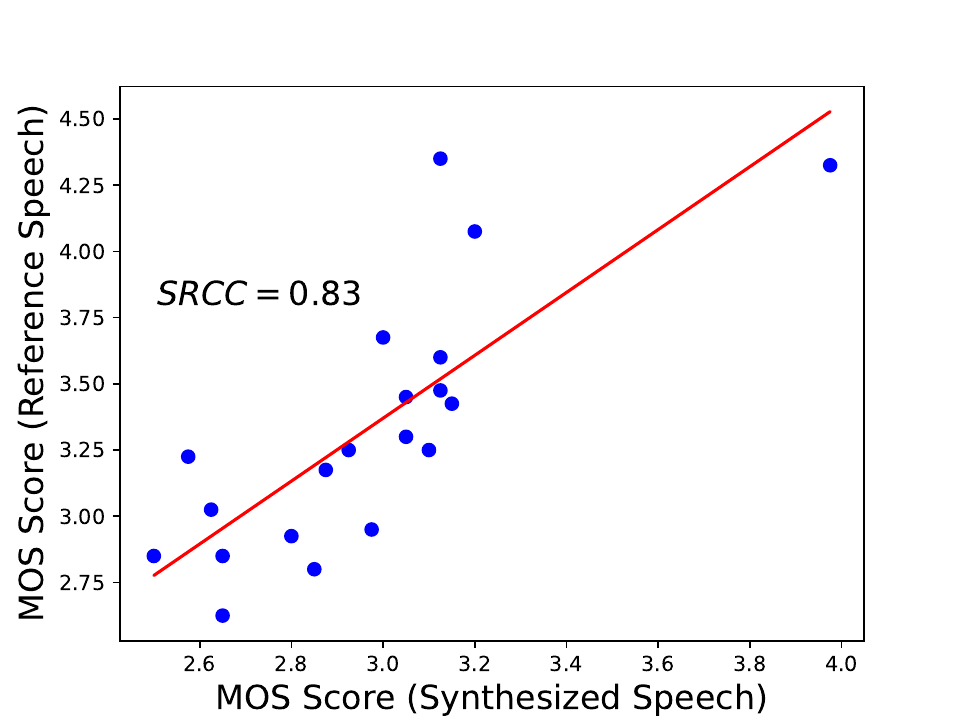}  
  \caption{Voice depth MOS score.}
  \label{fig:seen_voice_depth}
\end{subfigure}
\caption{\textbf{MOS score linear correlation visualization between synthesized speech and reference speech}. The Discriminative (w/ LoRA) system is used to generate speech for seen speaker scenario.}
\label{fig:visual_mos_score_corr}
\end{figure}


Results from the seen scenario indicate that, aside from the clarity attribute, our systems effectively capture the speaker's characteristics, with discriminative methods outperforming generative ones in terms of SRCC values. Despite this, as section \ref{ssec:fad_res} discusses, generative systems excel in creating high-fidelity audio. A synergistic approach, integrating both discriminative and generative techniques, achieves an optimal balance in preserving speaker characteristics and improving synthesized audio fidelity and naturalness. It should be noted that, apart from the pitch and speech attributes, which can be manipulated by signal processing strategy, our systems also capture the voice depth and age information from prompts very well. Manipulating these abstract concepts in speech is precisely the greatest strength of prompt-driven TTS systems. Besides, we also plot the MOS scores from synthesized and reference speech and visualize the linear correlation between them in Figure \ref{fig:visual_mos_score_corr}. The visualization further demonstrates that our system can capture the specific speaker characteristics from prompts.


Results from the bottom part of Table \ref{table:srcc_res} show that, for 
 the unseen speaker scenario, the system's ability to capture speaker characteristics in the prompt has weakened. This is because the prompt data amount in CSJ is still limited. In the future, we plan to train MOS predictors for speaker traits and use estimated MOS values for generating speaker impression prompts automatically for large amounts of speech data.

\section{Conclusion}
In this paper, we proposed to use prompts to specify and control the acoustic characteristics of the synthesized speech from a multi-speaker text-to-speech system. Different from previous works, listener impression scores are used to construct the prompts, thereby saving human resources and make the prompts closer to everyday expressions. Furthermore, we integrated a lightweight adapter module, LoRA, to efficiently fine-tune pre-trained language models for our specific requirements, yielding significant enhancements. Besides, we also decoupled the prompt-to-speaker module and the TTS system, which makes the whole system more flexible. To generate speaker embeddings from the prompt, we explored the discriminative method and flow-matching based generative method. Interestingly, We found that these two methods each have their own advantages, and combining them can further enhance the model.

\section{Acknowledgements}
This work was conducted during the first author's internship at NII, Japan. This study was partially supported by Google AI for Japan program. This work was partially supported in part by China NSFC projects under Grants 62122050 and 62071288, in part by Shanghai Municipal Science and Technology Commission Project under Grant 2021SHZDZX0102.

\bibliographystyle{IEEEtran}
\bibliography{mybib}

\newpage
\appendix
\section{Appendix}
Here, we evaluate our system from some other different perspectives.

\subsection{Speaker information relevance between synthesized speech and prompt according to speaker similarity}
Although we cannot consider the original speech in the evaluation set to be the ground truth of the speech generated by the corresponding prompt, the two should have a certain connection. For example, both should possess the speaker characteristics described in the prompt. Here, we generate speech by randomly selecting part of the original prompt in different proportions. We then assess the speaker similarity between the synthesized speech and its corresponding original speech. The findings, detailed in Table \ref{table:spk_sim_variance}, reveal a positive correlation between speaker similarity and the completeness of the prompt used for embedding generation, suggesting that our system effectively captures speaker-specific information. It is important to note that all systems demonstrated a cosine speaker similarity score below 0.5. This phenomenon stems from the fact that prompts do not encompass all characteristics of the target speakers and prompt-to-speaker task should be formulated as a one-to-many problem.

\begin{table}[ht!]
\footnotesize
\centering
\caption{\textbf{Speaker similarity variation when using different portions of the prompt on CJS evaluation set.}
}
\begin{adjustbox}{width=.43\textwidth,center}
\begin{threeparttable}
\begin{tabular}{l|ccc}
\toprule
\multirow{2}{*}{System} & \multicolumn{3}{c}{Portion of Prompt} \\
\cline{2-4} 
  & 1/3 & 2/3 & 3/3 \\
\hline
Discriminative (w/o LoRA) & 0.37 & 0.37 & 0.42 \\
Discriminative (w/ LoRA)&  0.36 & 0.38 & 0.41 \\
Flow-Matching (w/ LoRA) & 0.32 & 0.34 & 0.36  \\
Discriminative + Flow-Matching & 0.33 & 0.35 & 0.37 \\
\bottomrule
\end{tabular}
\end{threeparttable}
\label{table:spk_sim_variance}
\end{adjustbox}
\end{table}

\subsection{Speaker Embedding Visualization for Synthesized Audio from Different Prompts}
\begin{figure}[ht!]
\begin{subfigure}{.23\textwidth}
  \centering
  \includegraphics[width=1.0\linewidth]{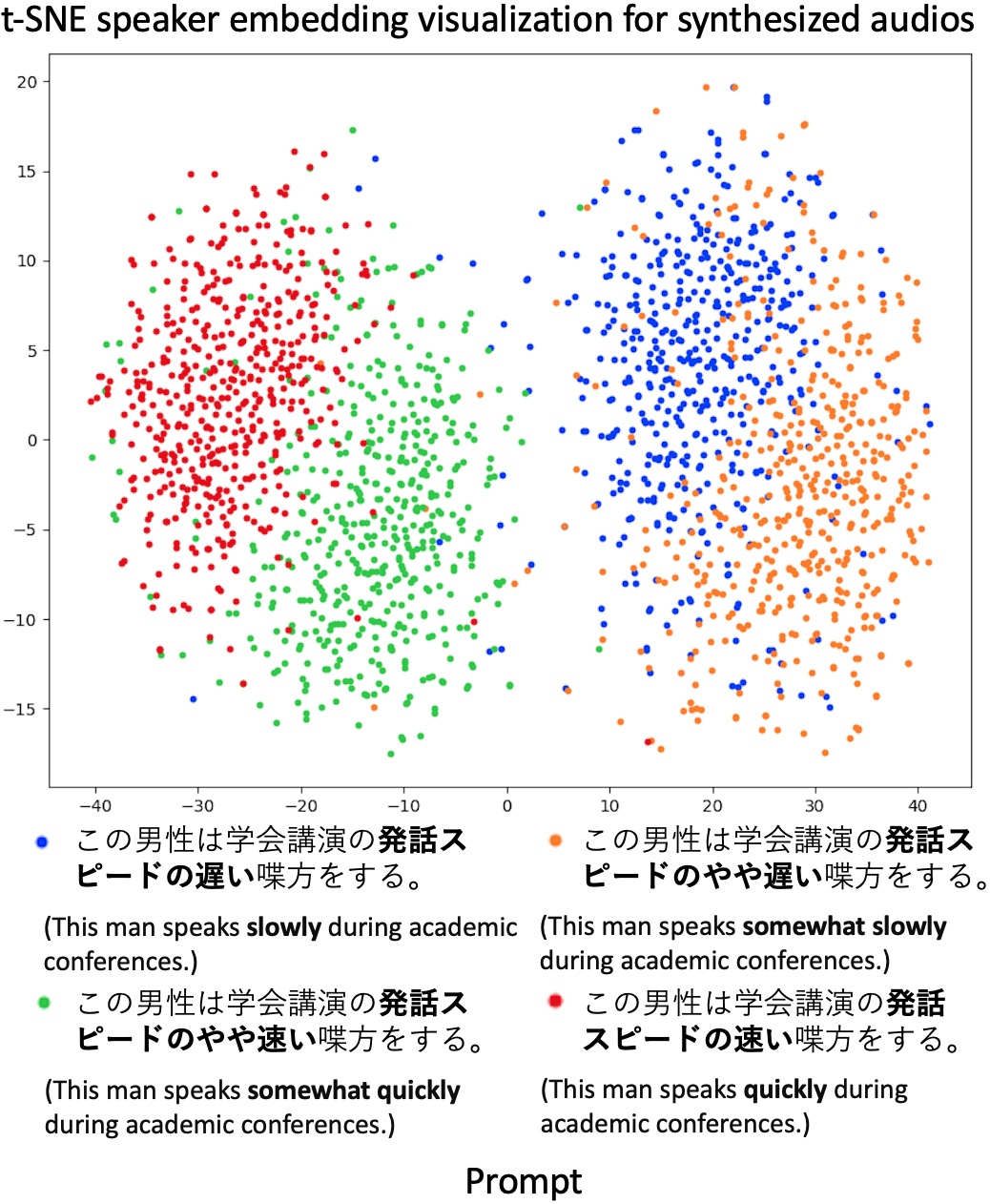}  
  \caption{Prompt related to speaking speed.}
  \label{fig:speed_visualize}
\end{subfigure}
\begin{subfigure}{.23\textwidth}
  \centering
  \includegraphics[width=1.0\linewidth]{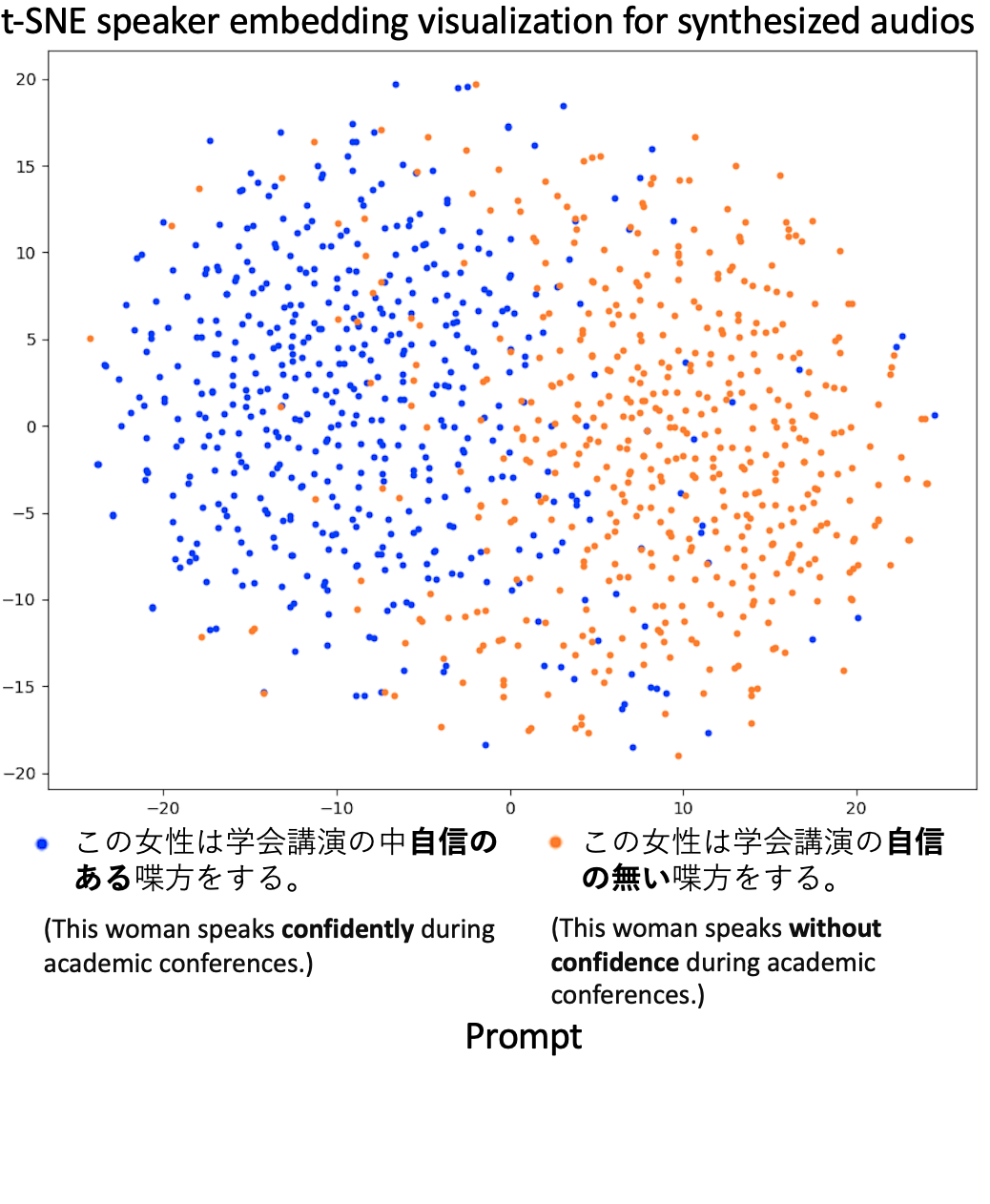}  
  \caption{Prompt related to speaking confidence.}
  \label{fig:confidence_visualize}
\end{subfigure}
\caption{t-SNE speaker embedding visualization for synthesized speech. Discriminative+Flow-Matching system is used in this figure. For each prompt, 500 audios are synthesized with the same context text.}
\label{fig:visualize}
\end{figure}

To further investigate whether some speaker-related attributes in the prompt truly have a controlling effect on the generated speech, we visualized the speaker embeddings extracted from the generated speech and plotted the visualization in Figure \ref{fig:visualize}. Figure \ref{fig:speed_visualize} illustrates that the speaker embeddings of synthetic voices from different prompts are clearly clustered into four classes, and the categories of prompts with similar concepts are even closer (e.g. the "slowly" cluster is close to "somewhat slowly" cluster). 
Additionally, the experiment underscores human language's unique capability to convey abstract concepts, such as the speaker's confidence level. The visualization in Figure \ref{fig:confidence_visualize} shows that the TTS system driven by prompts can effectively distinguish between concepts of different levels of speaking confidence.

\subsubsection{Synthesized Speeches' Attributes Distribution Visualization}
\begin{figure}[ht!]
\begin{subfigure}{.23\textwidth}
  \centering
  \includegraphics[width=1.0\linewidth]{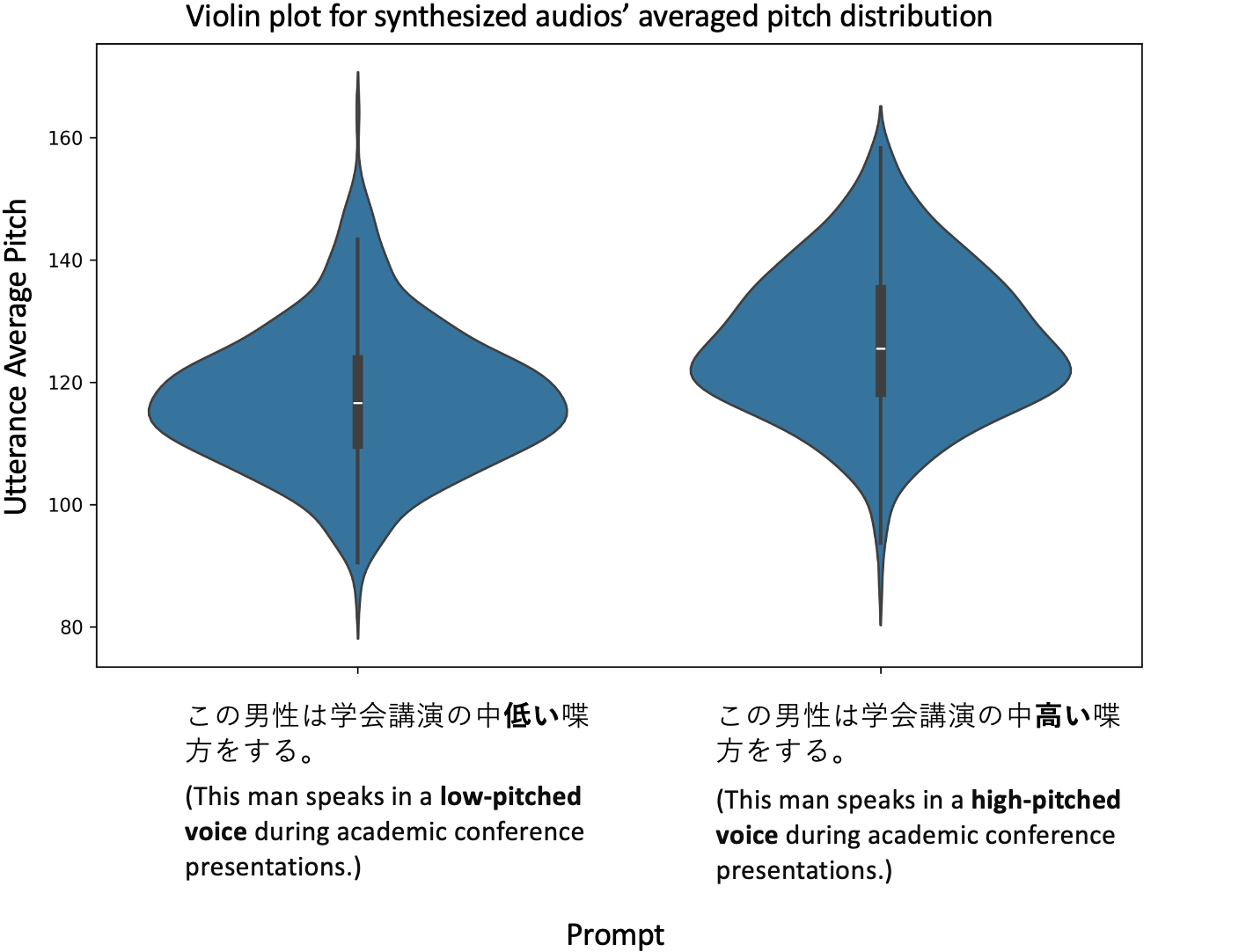}  
  \caption{Synthesized audios' averaged pitch distribution.}
  \label{fig:pitch_distribution}
\end{subfigure}
\begin{subfigure}{.23\textwidth}
  \centering
  \includegraphics[width=1.0\linewidth]{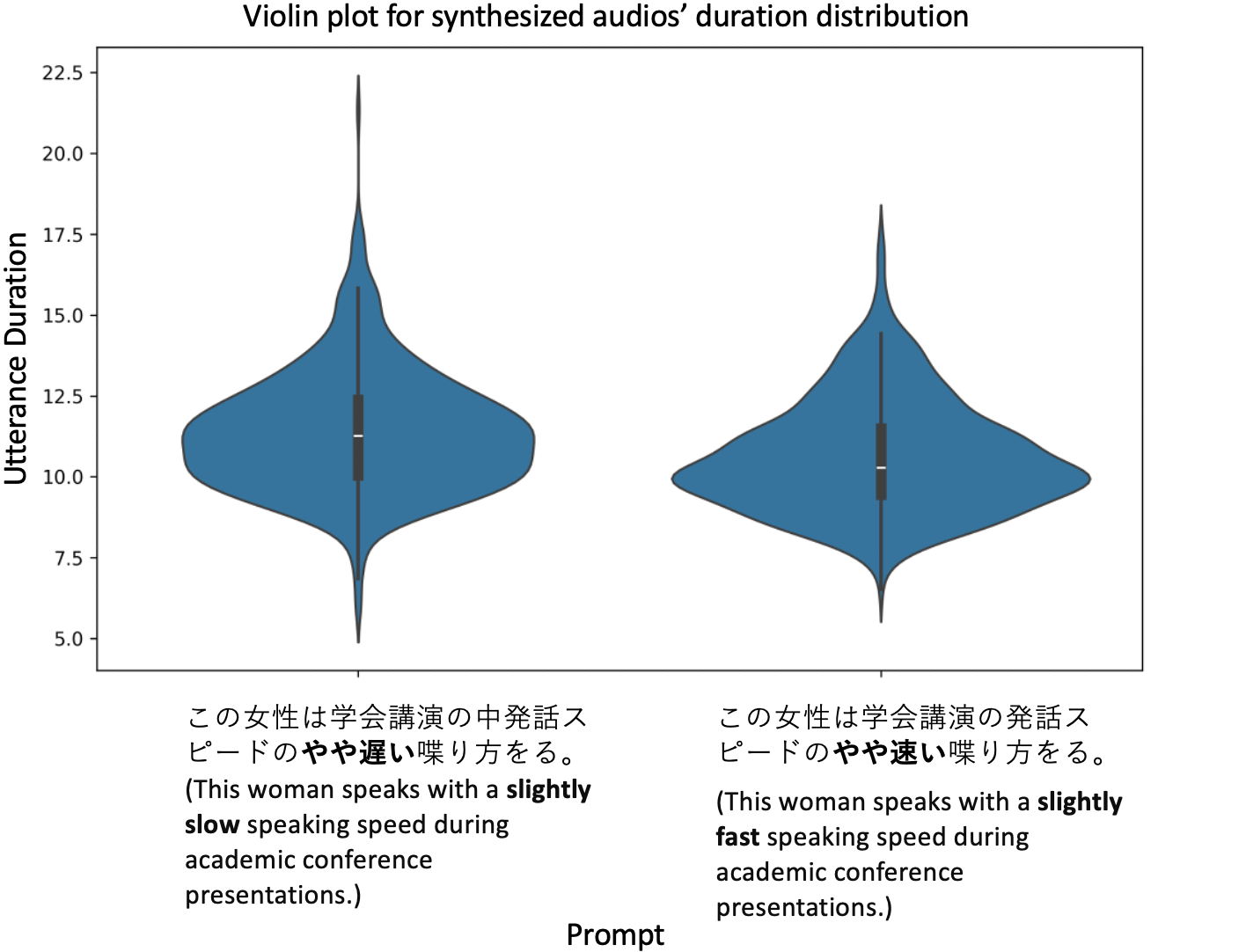}  
  \caption{Synthesized audios' duration distribution.}
  \label{fig:duration_distribution}
\end{subfigure}
\caption{Violin plot visualization for different attributes of synthesized speech. Discriminative+Flow-Matching system is used in this figure. For each prompt, 500 speeches is synthesized with the same context text.}
\label{fig:distribution}
\end{figure}

In this section, we present an analysis of the distribution of selected attributes for synthesized speeches generated from identical prompts. Because many descriptions in the prompt cannot be measured by objective indicators, we have selected the two attributes, pitch and speaking speed, to simply explore whether our system follows the description in the prompt. From Figure \ref{fig:pitch_distribution}, we can see that the speeches generated from the ``high-pitched" prompt do have an overall higher pitch distribution than the "low-pitched" one. Similarly, the distribution from Figure \ref{fig:duration_distribution} shows that the speeches generated from the ``slightly fast" prompt have an overall short duration. Different from the pitch and speaking speed information obtained from the signal processing measure in other work~\cite{guo2023prompttts,leng2023prompttts}, the CSJ dataset collects specific information based on the listener's subjective feeling. The results in this section further confirm that we can construct prompts from listener impression scores to control the speaker's characteristics in the speech synthesis task.

\begin{figure}[t!]
\begin{subfigure}{.5\textwidth}
  \centering
  \includegraphics[width=1.0\linewidth]{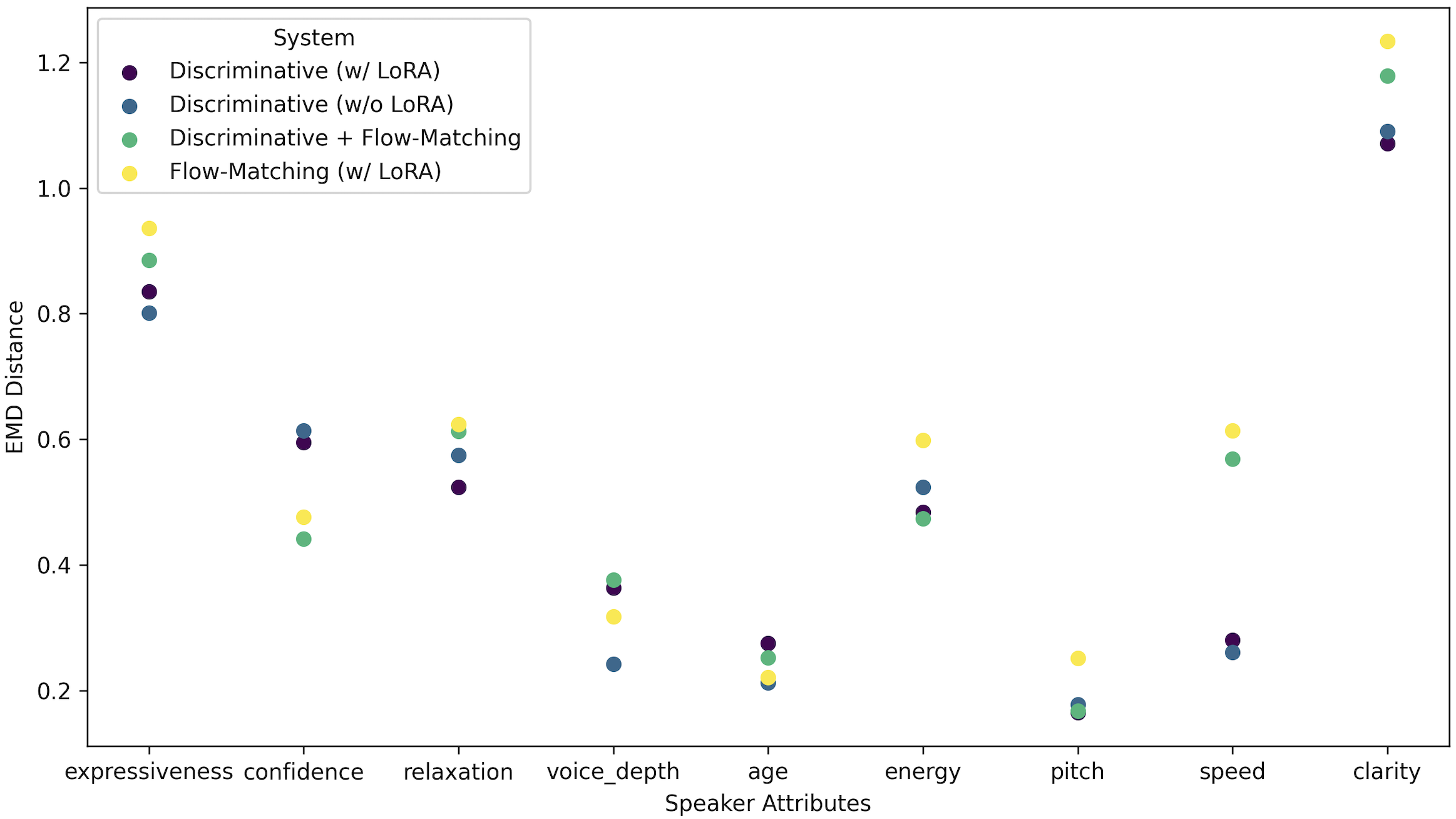}  
  \caption{Seen speaker scenario.}
  \label{fig:emd_seen}
\end{subfigure}
\begin{subfigure}{.5\textwidth}
  \centering
  \includegraphics[width=1.0\linewidth]{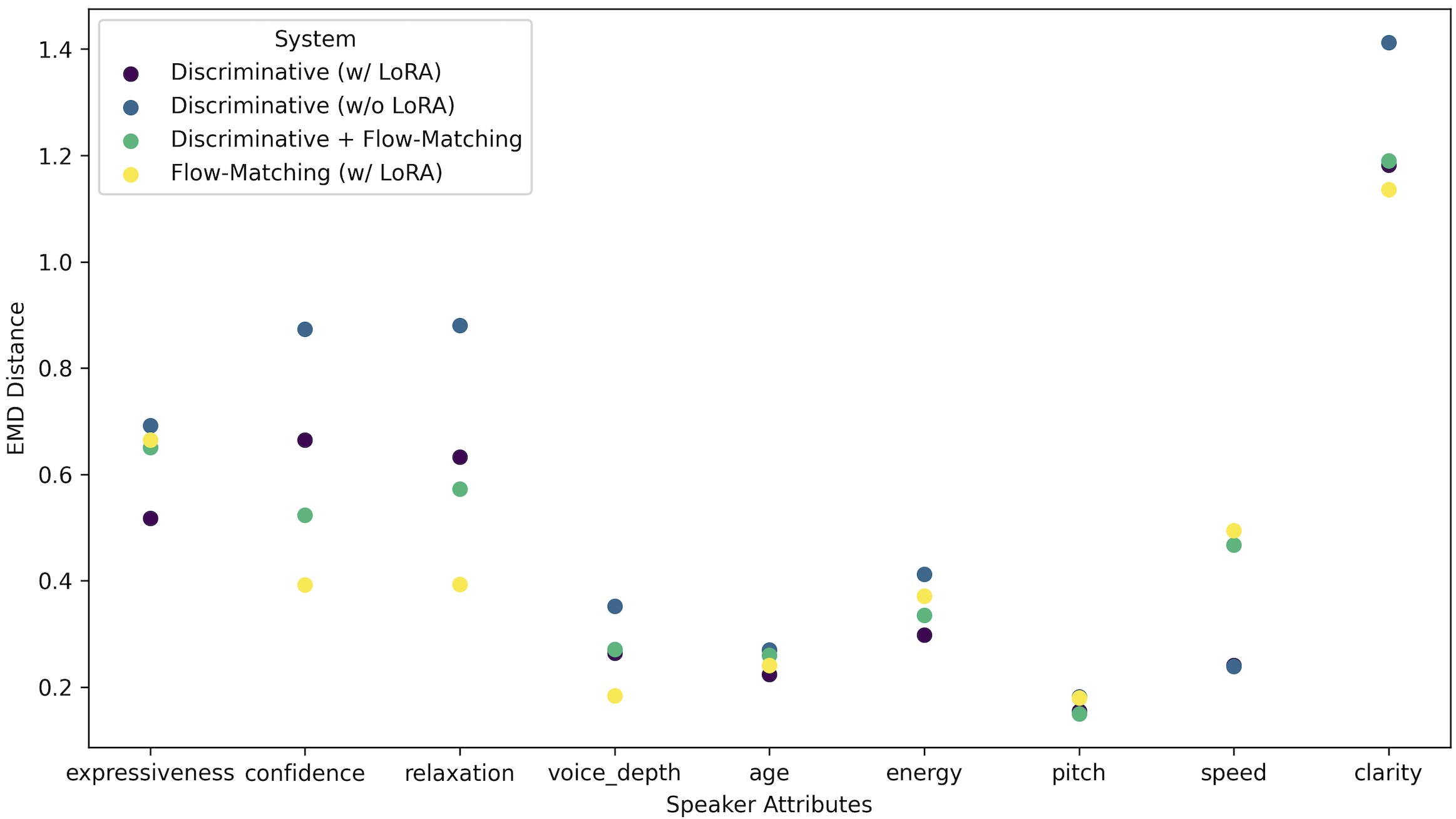}  
  \caption{Unseen speaker scenario.}
  \label{fig:emd_unseen}
\end{subfigure}
\caption{Earth Mover's Distance between MOS scores from synthesized speech and reference speech.}
\label{fig:emd}
\end{figure}

In section \ref{ssec:subjective_eval}, we average the MOS scores from the same speaker and attribute to one value. Here, we just leverage the original MOS scores and compute the Earth Mover's Distance (EMD) between MOS scores from synthesized speech and reference speech for each attribute. The results are shown in Figure \ref{fig:emd}. 
Unlike the SRCC results presented in Table \ref{table:srcc_res}, which quantify the alignment in variation trends of MOS scores between synthesized and reference speech, the EMD provides a measure of similarity in the numerical distribution of MOS scores between synthesized and reference speech. Essentially, the EMD assesses whether synthesized and reference speech share a comparable range in MOS scores.
The analysis revealed in Figure \ref{fig:emd} demonstrates that synthesized and reference speech exhibit closely matched MOS score scales across several attributes, including voice depth, age, energy, pitch, and speed.


\end{document}